# Study on the microwave ion source of 100-MeV proton linac

Hyeok-Jung Kwon

Korea Multipurpose Accelerator Complex, Korea Atomic Energy Research Institute
181 Miraero, Geoncheon-eup, Gyeongju-si, 38180, Republic of Korea
hjkwon@kaeri.re.kr

A microwave ion source is used as an ion source of 100-MeV proton accelerator at Korea Multipurpose Accelerator Complex (KOMAC). The specifications of the ion source are 50 keV in energy and 20 mA in peak current. The plasma is operated in CW mode using magnetron and the pulse beam is extracted using semiconductor switch located in the extraction power supply. The beam characteristics were measured based on the pulse voltage and current. A test stand was also installed to study the beam characteristics of the ion source in off-line. In this paper, the pulse beam characteristics of the ion source are presented and the installation of the test stand is reported.

# I. INTRODUCTION

A microwave ion source is used for 100-MeV proton accelerator at KOMAC. The operating parameters are summarized in Table 1 and the geometry of the microwave ion source is shown in Fig. 1. The microwave components of the ion source are a magnetron, directional coupler, 3-stubs tuner, isolation waveguide, ridge waveguide and microwave window. It uses a 2.45 GHz, 1 kW magnetron as a microwave source. An isolation waveguide consists of 30 sets of aluminum plates and G-10 plates to isolate the magnetron from the high-voltage potential of the ion source. The boron nitride and aluminum nitride plates are used as a microwave window. A single solenoid magnet is installed around the arc chamber to confine the plasma. The extraction geometry has three electrodes configuration: one is a plasma electrode, another is a bias electrode and the third is a ground electrode as shown in Fig. 1. During operation, the plasma is always turned on and the extraction power supply is switched on and off using a semiconductor switches. The pulse beam characteristics were investigated to understand the pulse mode operation of the ion source using the semiconductor switch consisted of MOSFET molded with the silicon delivered from a foreign company called BELKE [1]. But we replaced that switch with the oil type switch developed in domestic company due to the easiness of the maintenance.

In this paper, we want to focus the pulse beam characteristics of the ion source with new semiconductor switch. The circuit of the extraction power supply is described in Section II, the results of the pulse beam measurement are described in Section III, installation of the test stand of the ion source is described in Section IV and the conclusion is presented in Section V.

# II. BEAM EXTRACTION SYSTEM

The circuit diagram of the pulse beam extraction system is shown in Fig. 2. A high voltage power supply is used to charge the energy storage capacitor. A semiconductor switch is installed after the capacitor. The switch is a push pull type in order to put the potential of the ion source to ground during off state. Total 80 sets of IGBT switches are connected in series and installed inside oil for high voltage insulation. A 20 kohm resistor is installed in the oil tank of the switch to limit the maximum current when there is an arc in the load. There is another high voltage power supply system for a bias electrode. It also uses the same type of switch with low rated voltage. The characteristics of the switch itself were tested. The results are shown in Fig. 3 and Fig. 4. The rising time was 10us and falling time 20 us.

### III. PUSLE BEAM MEASUREMENT

The pulse beam from the ion source of the 100-MeV proton linac was measured. A high voltage probe was used to measure the extraction voltage. An AC Current Transformer (ACCT) located in front of the RFQ was used to measure the beam current. Also a pulse signal to trigger the switch was measured. The measured signal is shown in Fig. 5. The measured conditions are such that a pulse width was 2 ms, an extraction voltage was 50 kV, bias voltage was 3 kV, beam current was 15 mA. The delay between applied voltage and extracted current was less than 5 us and the rising time of the beam current was 5us as shown in Fig. 6. The voltage droop was 1 kV during 2 ms and it induces a 1% beam current droop as shown in Fig. 7. When we consider the capacitance of the energy stored capacitor, beam current and pulse width, it droop is well agreed with the calculated one. The long term stability of the beam current at the ion source was also measured and the result is shown in Fig. 8. The standard deviation of the pulse to pulse beam current was 0.2%.

### IV. TEST STAND OF THE ION SOURCE

There is a limitation to use an ion source of the 100-MeV proton linac directly to improve the performance of the ion source itself because it is a part of the machine for user service. Therefore, a test-stand of the microwave ion source was developed independently. It has same geometry but have some improvements. First it used an alumina as an insulator instead of the Teflon which was used for ion source of the 100-MeV linac. Second it isolated the solenoid magnet from the high voltage side using the Nylon insulator as shown in Fig. 9. As a consequence, it can remove all power supplies located at high voltage side and isolation transformer. Recently, the high voltage test was done up to 55 kV and the pulse beam extraction was carried out.

## V. CONCLUSION

The pulse beam characteristics were measured from the microwave ion source for 100-MeV proton linac. The rising time, droop and stability were met with the operation conditions of the linac. But we could measure the current ripple which has the frequency of 20 kHz, which is the same frequency of the modulator IGBT switching frequency. We are going to investigate the effect of the ripple further. In addition, it is necessary to improve the ion source continuously. Therefore, test stand was developed. Not only the beam characteristic measurements but also the hardware improvements are planned at the test stand.

## ACKNOWLEDGMENTS

The author would like to thank to Dr. Han-Sung Kim, Mr. Dae-Il Kim, Mr. Dong-Hyeok Seo and Dr. Yong-Sub Cho for their discussion of and valuable comments on this work. This work was supported by the Ministry of Science, ICT & future Planning of the Korean government.

Table 1. Operating parameters of the microwave ion source.

| Parameters | Values |
|---|---|
| Particle | Proton |
| Beam energy | 50 keV |
| Maximum peak beam current | 20 mA |
| Emittance (normalized rms) | 0.2 π mm mrad |
| Proton fraction | > 80 % |
| Operation time without maintenance | > 100 hrs |
| Microwave frequency | 2.45 GHz |

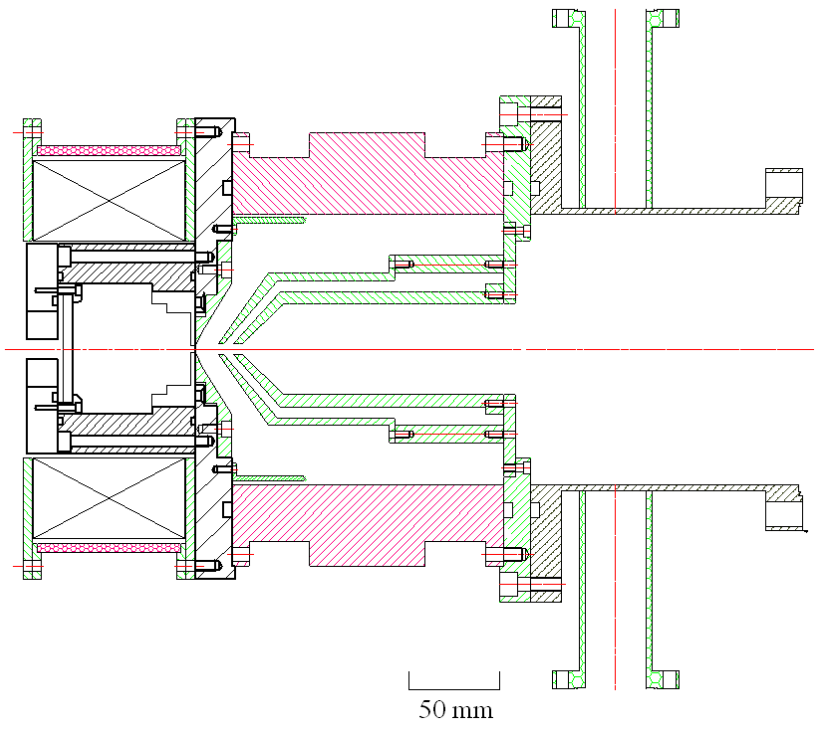

Fig. 1. Geometry of the microwave ion source.

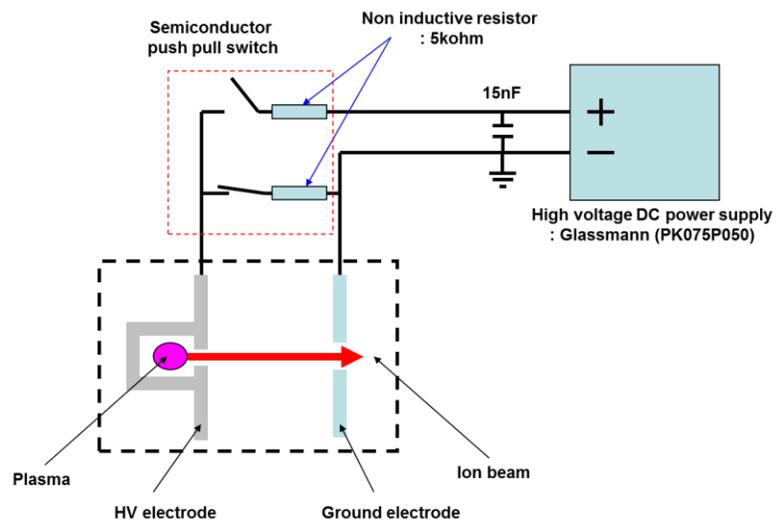

Fig. 2. Circuit diagram of the high voltage switch.

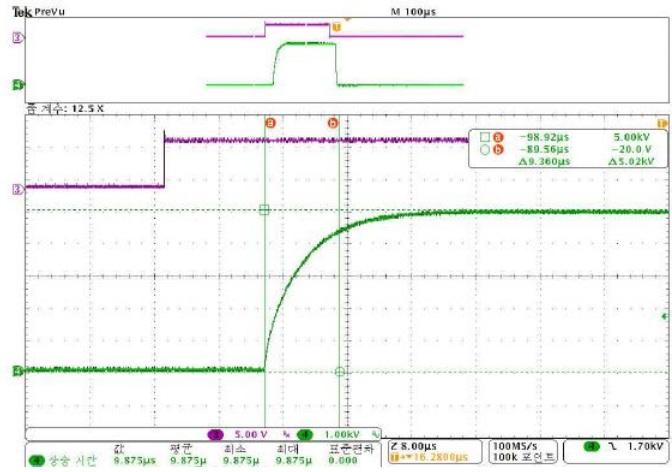

Fig. 3: Rising edge of the pulse using switch with resistive load (Green).

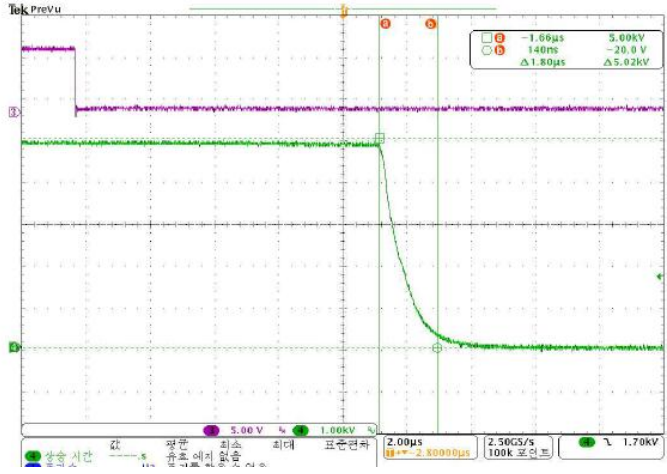

Fig. 4: Falling edge of the pulse using switch with resistive load (Green).

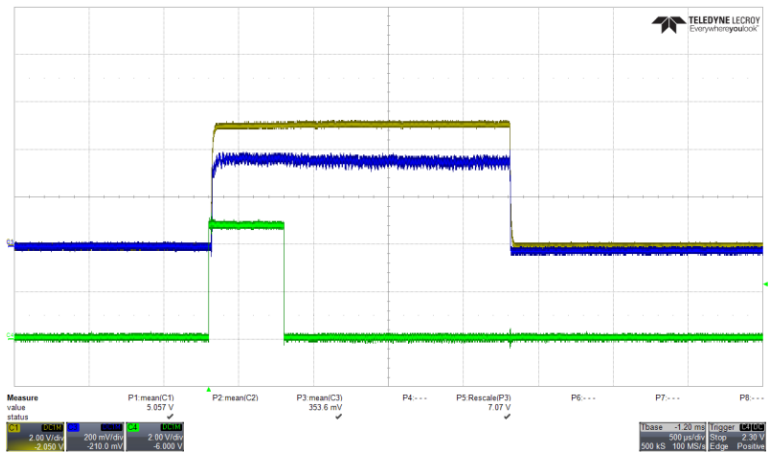

Fig 5: Pulse beam shape (Ch1: extraction voltage, Ch2: beam current, Ch4: trigger)

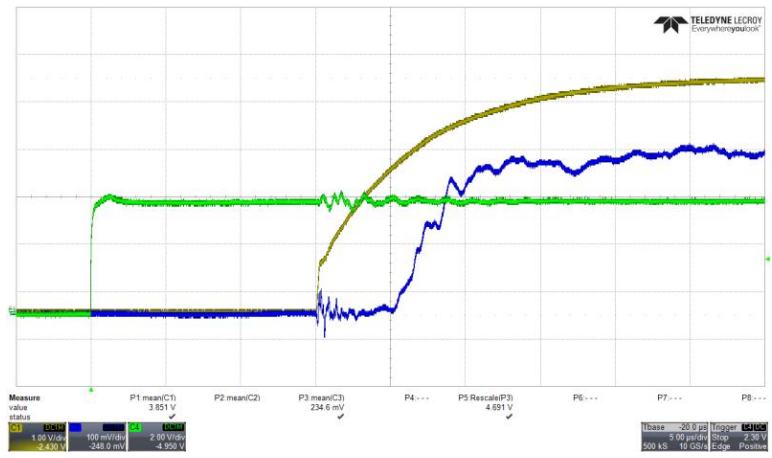

Fig. 6. Rising edge of beam (Ch1: extraction voltage, Ch2: beam current, Ch4: trigger)

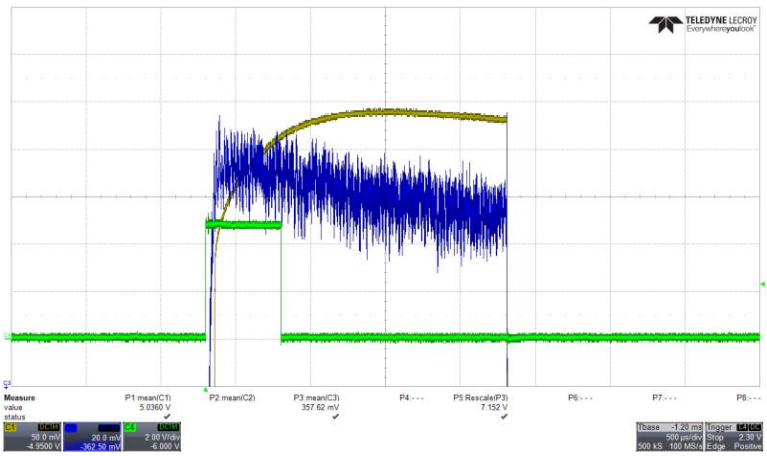

Fig. 7. Beam current droop (Ch1: extraction voltage, Ch2: beam current, Ch4: trigger)

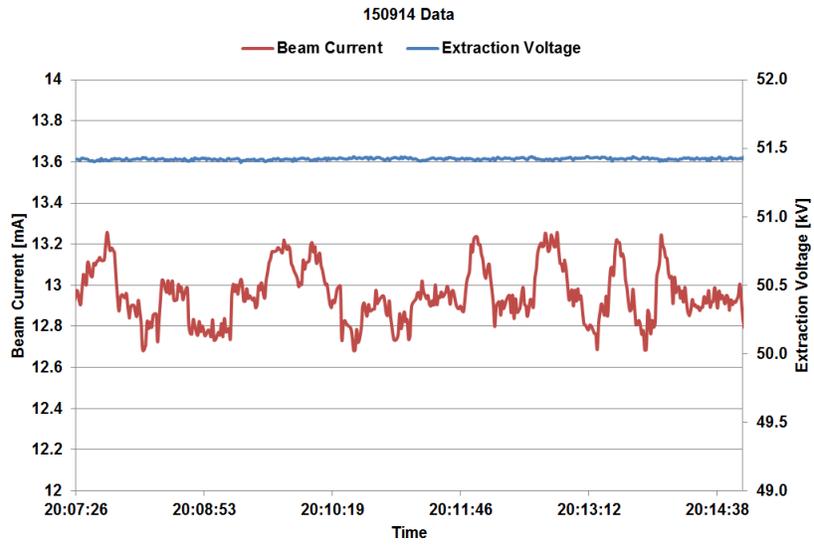

Fig. 8. Pulse to pulse stability measurement.

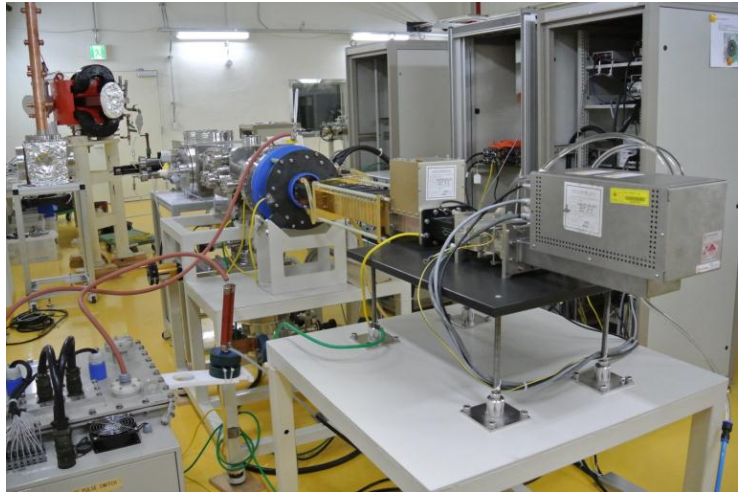

Fig. 9. Microwave ion source test stand